\documentclass[12pt]{iopart}
\usepackage{epsf}
\begin{document}

\title[Non-identical correlations in STAR]{Non-identical particle correlations
in 130 and 200 AGeV collisions at STAR }

\author{Adam Kisiel (for the STAR Collaboration\footnote[1]{For the full author list and acknowledgements see Appendix "Collaborations" in this volume.})}

\address{Faculty of Physics, Warsaw University of Technology, Koszykowa 75, 00-662 Warszawa, Poland}

\ead{kisiel@if.pw.edu.pl}

\begin{abstract}
STAR has performed a correlation analyses of pion-kaon and pion-proton pairs for $\sqrt{s_{NN}}=130 AGeV$ and $\sqrt{s_{NN}}=200 AGeV$ and kaon-proton, proton-Lambda and pion-Cascade pairs for AuAu collisions $\sqrt{s_{NN}}=200 AGeV$. They show that average emission space-time points of pions, kaons and protons are not the same. These asymmetries are interpreted as a consequence of transverse radial expansion of the system; emission time differences explain only part of the asymmetry. Therefore our measurements independently confirm the existence of transverse radial flow. Furthermore, correlations of strange hyperons is investigated by performing proton-Lambda and pion-Cascade analyses, giving estimates of source size at high $m_{T}$. The strong interaction potential between (anti-)proton and lambda as well as kaon and proton is investigated.
\end{abstract}




One of the main areas of interest in studying relativistic heavy-ion collisions is the collective behavior of matter, or flow. Non-identical particle correlations ~\cite{ledn,ledn1,xupan} have been proposed~\cite{lednall} as a measure of the transverse radial flow and asymmetries in particle emission from the expanding system produced in such collisions.

\section{Non-identical particle correlations technique}
Identical particle correlations (HBT) have been used to study the space-time characteristics of the in heavy-ion collisions~\cite{pipihbt}. Non-identical particle correlations give new insights into collision's dynamics~\cite{lednall}. Such correlations arise from coulomb (for charged particles) and strong (for hadrons) interactions, which occur after particles' last elastic collision. We show correlation functions for particle pairs with small relative momentum $k^{*}$, which means close velocities in the source rest frame. The height and width of the correlation effect carry the information about the size of the emission region. 
\par The correlation functions of non-identical particles also carry a qualitatively new piece of information - the emission asymmetries~\cite{ledn}. The technique uses the fact that both particles are identifiable. We divide particle pairs in two groups - with $k^{*}_x > 0$ and $k^{*}_x < 0$, where $k^{*}_x$ is the component of the relative momentum of the first particle in one of the following directions: $out$ - the direction of pair momentum, $long$ - the direction of beam axis or $side$ - perpendicular to both previous ones. We can then construct two correlation functions: $C^{x}_{+}(k^{*})$ for pairs with $k^{*}_x > 0$, and $C^{x}_{-}(k^{*})$ for pairs with $k^{*}_x < 0$. It can be shown that these two functions should be identical if the average emission points of the two particle species are the same. However if there is a non-zero difference $\langle \Delta r_{x} \rangle$ between average emission points $\langle r_{x} \rangle$ of both particle species in the given direction $x$, the functions $C^{x}_{+}$ and  $C^{x}_{-}$ will be different, and their difference, visible in the ``double ratio'' $C^{x}_{+}(k^{*})/C^{x}_{-}(k^{*})$ will be directly related to this $\langle \Delta r_{x} \rangle$. 
\par By symmetry considerations it can be shown, that $\left\langle \Delta r_{side} \right\rangle = 0 $. For a symmetric system with a symmetric rapidity coverage (as is the case in AuAu collisions at STAR) we also have $\left\langle \Delta r_{long} \right\rangle = 0 $. The only direction where we expect the asymmetry is $out$. It can be written:
\begin{equation} 
\left\langle r^{*}_{out}\right\rangle = 
\left\langle\gamma (\left\langle\Delta r_{out}\right\rangle - \beta \left\langle\Delta t\right\rangle)\right\rangle
\label{rousdep}
\end{equation}, 
where values with asterisk ($^{*}$) are in the pair rest frame and therefore can be measured directly through the correlation functions, while values without the asterisk are in the source rest frame. From equation (\ref{rousdep}) one can see that the observable asymmetry can come from space and/or time component. 

\section{Results from the STAR experiment}
The non-identical particle (pion-kaon) correlations have been studied by the STAR experiment~\cite{pikcorrel}. In a continuation of this analysis we have used pions, kaons and protons identified in the STAR Time Projection Chamber (TPC) detector~\cite{tpc}. Central AuAu collisions at $\sqrt{s_{NN}}=130 AGeV$ and $200 AGeV$ were studied. Particles with rapidity ($-0.5 < y < 0.5$) were used. The momentum ranges (in $GeV$) were: (0.08, 0.5) for pions, (0.3,1.0) for kaons and (0.3,1.2) for protons. All functions were corrected for particle purity, momentum resolution and two-track detector effects (hit-sharing) - for detailed description we refer the reader to~\cite{pikcorrel}.

\par We have measured correlations with strange baryons - $\Lambda$ and $\Xi$. They were identified using a topological reconstruction of their decay into $\pi$ and $p$. Particle purity has been estimated accounting for resonance feed-down and the signal to noise under the invariant mass peak. The momentum range for $\Lambda$'s was $0.3-2.0$ $GeV/c$. 

\begin{figure}
  \begin{center}
      \epsfxsize=100mm
      \epsfysize=60mm
      \epsfbox{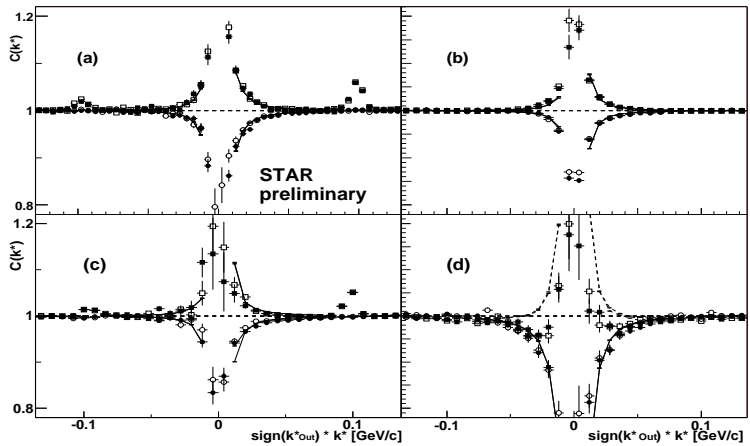}
      \vspace{-0.5cm}
\caption{The correlation functions for (a) pion-proton at $130 AGeV$, (b) pion-kaon at $200 AGeV$, (c) pion-proton at $200 AGeV$ and (d) kaon-proton at $200 AGeV$. The \fullcircle points are for $++$ combination, \opencircle for $--$, \fullsquare~for $+-$ and \opensquare~for $-+$. Solid lines show fits to the data, dashed lines show theoretical expectation.\label{cffig}}
  \end{center}
\vspace{-0.6cm}
\end{figure}


\par In Fig. \ref{cffig} we present the preliminary correlation functions for all combinations of pion-kaon, pion-proton kaon-proton pairs. The correlation functions for all like-sign and opposite-sign combinations agree, which suggests that the emission mechanisms for opposite charge pions, kaons and protons are the same. It is important to note that opposite-charge kaon-proton correlation function clearly deviates from the theoretical expectation, which suggests that our knowledge of this interaction is incomplete. 

\par We obtain the ``double'' ratios in the $out$ direction by dividing the right side ($k*_{out}>0$) of the plots in Fig. \ref{cffig} by the left($k*_{out}<0$). They clearly deviate from unity for all the pair combinations, showing that pions, kaons and protons are not emitted from the same average space-time point. 

\par To quantify that shift a fitting procedure is applied to the data. A source is assumed to be a 3-dimensional Gaussian in the pair rest frame with a width $\sigma$ and a shift $\langle r^{*}_{out} \rangle$ in the $out$ direction. They are taken as parameters of the fit. For each of their values, with the use of the experimental momentum distribution, a theoretical correlation function is constructed. We use a $\chi^{2}$ test to measure how well the generated correlation function agrees with the experimental one. The one with the lowest $\chi^{2}$ value is taken as the ``best fit''. The results of this fitting procedure are shown on Fig. \ref{modelfit}. 

\section{Understanding space-time asymmetries from models}
Heavy-ion reaction models, which study the dynamics of the collision, provide a description of the emitting source. They also give the predictions for the emission asymmetries, which can be directly compared to our experimental results and used to gain a better understanding of the reaction.

\begin{figure}
\begin{center}
\epsfxsize=130mm
\epsfbox{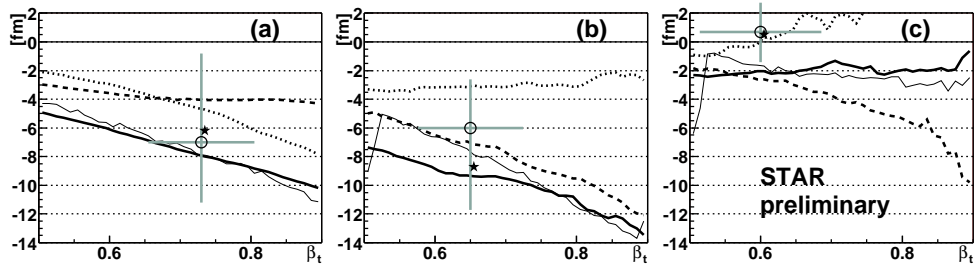}
\vspace{-0.2cm}
\caption{Comparing (a) pion-kaon, (b) pion-proton and (c) kaon-proton correlation function fit results (\opencircle symbols) for $\langle r^{*}_{out} \rangle$ with model predictions: blast-wave parameterization(thin solid line) and RQMD(thick solid line). We show a space(dotted line) - $ \langle \Delta r_{out} \rangle$ and time (dashed line) - $\langle \Delta t \rangle$ component of the asymmetry from RQMD separately. The results of the fit to the RQMD simulations are shown as $\star$ symbols.}
\label{modelfit}
\end{center}
\vspace{-1.0cm}
\end{figure}

\par
The blast-wave parameterization~\cite{blast} inspired by hydrodynamic calculations provides a simple way of describing a system with strong transverse radial flow. The asymmetry comes from the interplay between a flow velocity and thermal smearing of the emission point. This model shows that radial flow does produce the emission asymmetry that we expect. Moreover, the blast-wave calculation~\cite{blast} done with the parameters fixed by other measurements (elliptic flow - $v_{2}$~\cite{v2}, $\pi\pi$ HBT~\cite{pipihbt}, single particle spectra~\cite{spectra}) predict an asymmetry which is consistent with the data~\cite{pikcorrel}. 
\par
We also studied the predictions of the RQMD model~\cite{rqmd}, which produces radial flow through hadronic rescatterings. In addition it allows for a complete treatment of resonances. Their decay can be a source of delayed (relative to the direct production) emission of pions, kaons and protons. Since the average delays are different for different  particle species, we see a resulting average time shift between pions, kaons and protons, which produces emission asymmetries which sum with the spatial asymmetry coming from flow (see eq. (\ref{rousdep})). The relative weight of both components is shown on Fig. \ref{modelfit}. Experimentally, it is impossible to disentangle space and time contributions to the asymmetry. However the study of rescattering models shows, that neither flow, nor time difference alone are able to explain the magnitude of asymmetries observed in the data. This suggests that both are necessary to explain our measurements.

\section{Correlations with strange baryons}

\begin{figure}
\begin{center}
\epsfxsize=140mm
\epsfbox{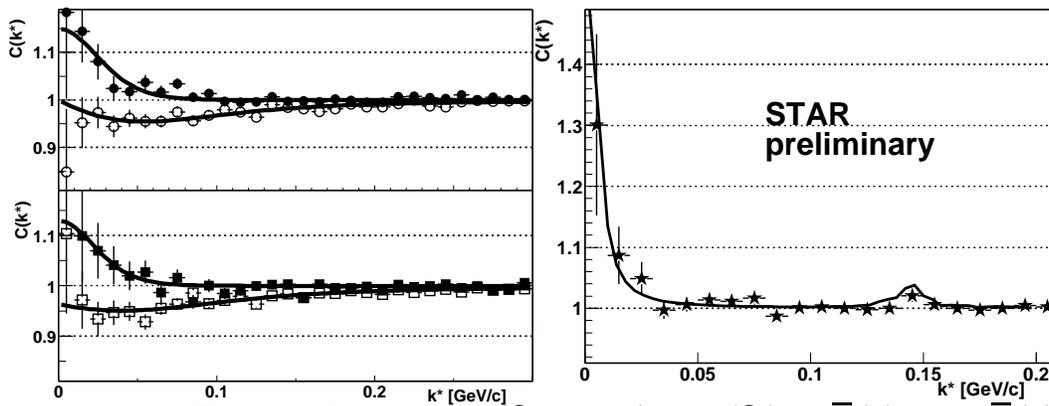}
\vspace{-0.5cm}
\caption{Left panel: $p\Lambda$ ($\fullcircle$ symbols), $\overline{p}-\Lambda$ ($\opencircle$), $p-\overline{\Lambda}$ ($\fullsquare$) and $\overline{p}-\overline{\Lambda}$ ($\opensquare$) correlation functions from STAR. Solid line: fit to the data. Right panel: Raw combined $\pi^{+}\Xi$ and $\pi^{-}\overline{\Xi}$ correlation functions ($\star$) from STAR at $200 AGeV$, solid line: theoretical expectation.}
\label{placf}
\end{center}
\vspace{-0.8cm}
\end{figure}

The $p\Lambda$ correlation function from STAR~\cite{gaelwar} is presented in Fig.~\ref{placf}. It has been studied using the analytical model from Lednick\'{y} and Lyuboshitz~\cite{ledn1}. If we assume a Gaussian source, and use the interaction parameters from~\cite{prattpl}, we obtain a source size $r_0 = 2.72 \pm 0.33 fm$. An intercept parameter $\lambda$ was fixed at $0.15$, which is consistent with estimated pair purity. In Fig.~\ref{placf} we also show, measured for the first time, $\overline{p}-\Lambda$ correlation function. For this system the phase shifts are not known. The same model as for $p\Lambda$ has been applied, but this time with phase shifts as free parameters. The radius $r_0 = 1.42 \pm 0.07 fm$ was obtained, as well as a non-zero imaginary part of the scattering amplitude, which suggests significant annihilation in the system. 

\par
In Fig.~\ref{placf} we present the preliminary $\pi\Xi$ correlation function from STAR. It has a shape expected for a Coulomb-dominated interaction. The $\Xi^{*}(1530)$ resonance is observed at $k^{*} = 145 MeV/c$. The calculation shown was performed using Pratt's $\pi\Xi$ interaction calculation code, with the assumption that the $\Xi$ flows significantly. In the future this function may provide new insights into the question of how does the $\Xi$ flow compared to pions.

\section{Conclusions}
We have presented the preliminary results of pion-kaon, pion-proton and kaon-proton correlations measurements by the STAR experiment. The emission asymmetries between all types of particles have been measured. They are shown to be consistent with transverse radial flow produced in heavy-ion collisions, as seen e.g in the blast-wave parameterization. Rescattering models were used to study the effect of emission time differences, which were found to be important, but not sufficient to explain the observed asymmetry. Therefore our measurements provide an independent confirmation of the existence transverse radial flow in AuAu collisions at RHIC. The $\pi\Xi$ correlations are also supporting this picture; such measurement may provide, in future, important information on collective behavior of $\Xi$.
\par Correlations with strange baryons (as well as opposite-charge kaon-proton correlations) have been used to study the unknown interaction potentials, the information which is sometimes inaccessible in any other way (e.g. $\overline{p}\Lambda$ correlations). They have also been used to obtain source sizes at high $m_T$.

\end{document}